\documentclass[twocolumn,amsmath,amssymb,prl]{revtex4}
\usepackage{graphicx}% Include figure files
\usepackage{dcolumn}% Align table columns on decimal point
\usepackage{bm}% bold math

\def\beq{\begin{equation}}
\def\eeq{\end{equation}}

\begin{document}
\input{epsf}

\title{The Race Between Stars and Quasars in Reionizing Cosmic Hydrogen}

\author{Abraham Loeb}

\affiliation{Astronomy Department, Harvard University, 60 Garden St.,
Cambridge, MA 02138, USA}

\begin{abstract} 

The cosmological background of ionizing radiation has been dominated
by quasars once the Universe aged by $\sim 2$ billion years. At
earlier times (redshifts $z\gtrsim 3$), the observed abundance of
bright quasars declines sharply, implying that cosmic hydrogen was
reionized by stars instead. Here, we explain the physical origin of
the transition between the dominance of stars and quasars as a generic
feature of structure formation in the concordance $\Lambda$CDM
cosmology.  At early times, the fraction of baryons in galaxies grows
faster than the maximum (Eddington-limited) growth rate possible for
quasars. As a result, quasars were not able to catch up with the rapid
early growth of stellar mass in their host galaxies.

\end{abstract}

%\paragraph*{Keywords:} Galaxies: Massive Black Holes

%\date{\today}
\maketitle

\paragraph{Introduction.}
Observations of the Ly$\alpha$ forest and the quasar luminosity
function at redshifts $z\lesssim 6$, suggest that quasars dominated
the production rate of hydrogen-ionizing photons only after the
Universe has aged by 1-2 billion years \citep[see Fig. 5 in
Ref.][]{Claude08}. The sharp decline in the observed comoving density
of bright quasars at redshifts $z\gtrsim 3$ implies that supermassive
black holes (BHs) could not have produced sufficient UV photons to
reionize cosmic hydrogen by a redshift of $z_{\rm reion}=10.9\pm 1.4$,
as required by the WMAP5 data on the microwave background anisotropies
\cite{Komatsu}.  It is therefore widely believed that stars have
reionized cosmic hydrogen \cite{Claude08,UNESCU}. Here we suggest a
simple physical explanation for this phenomenological inference.

\paragraph{Preliminaries.}
Quasars are powered by the accretion of gas onto massive BHs
\cite{Salpeter,Acc}. The remnant BHs they leave behind are observed in
the nuclei of all bulged galaxies at the present time \cite{Lauer}. BH
growth is accelerated during episodes of galaxy mergers, when cold gas
is infused to galactic nuclei by tides; this association is predicted
by computer simulations \cite{Cox,Springel,Li} and supported by
observations \cite{Merg_Obs}. The observed correlation between the
mass of nuclear BHs and the depth of the gravitational potential well
of their host galactic bulges (as inferred from the velocity
dispersion of stars) \cite{Geb}, suggests that their growth was
ultimately self-regulated \cite{Silk,WL03}.  Merely $\sim 5\%$ of the
energy output from a bright quasar is sufficient to unbind the cold
gas reservoir from its host galaxy \cite{WL03,Springel}.

Bright quasars are inferred to be radiating close to their limiting
{\it Eddington} \cite{Edd} luminosity, $L_E$.  Estimates of their BH
masses imply typical luminosities $L$ in the range of $\sim
(0.1$--$1)\times L_E$ \citep[see Fig. 6 in Ref.][]{Kol}. The
Eddington limit is derived by equating the repulsive radiation force
on an ionized gas element to the attractive gravitational force on it
towards the BH \cite{SL},
\begin{equation}
L_E \equiv {4\pi c Gm_p M\over \sigma_T}= 1.4\times 10^{44}~{\rm
erg~s^{-1}} \left({M\over 10^6 M_\odot}\right) ,
\end{equation}
%Avi: multiply by 1.14=1/0.88 for primordial composition 
%and doubly-ionized helium. This increases the value of t_E by 14%
%which exactly compensates for the Sheth-Tormen modification. But
%avoided here for simplicity.
where $\sigma_T$ is the Thomson cross-section for electron scattering
and $M$ is the BH mass.  In a spherical geometry, the radiation force
associated with $L>L_E$ would generate an outflow and inhibit
accretion onto the BH.  Given an unlimited fuel reservoir, the growth
rate of a BH would be regulated by the maximum luminosity that
allows it to accrete gas, $L_E$.

The luminosity of a quasar is related to the gas accretion rate
$\dot{M}_{\rm acc}$ through $L=\epsilon\dot{M}_{\rm acc}c^2$, where
$\epsilon$ is the efficiency for converting the rest mass of the
accreting gas to radiation.  At high accretion rates, the BH
luminosity is expected to approach its limiting value $L_E$
\cite{Narayan}, as inferred for bright quasars \cite{Kol}.
Theoretical models imply that at $L\gtrsim 0.5L_E$, the accretion disk
is puffed-up by radiation pressure \cite{Shafee1,Shafee} and its
geometry enters the quasi-spherical regime for which the Eddington
limit is derived.

The growth rate of the BH mass, $\dot{M}=(1-\epsilon)\dot{M}_{\rm
acc}$ is related to the {\it Salpeter} \cite{Salpeter} time,
\begin{equation}
t_E\equiv {M\over \dot{M}}= 4\times 10^8~{\rm yr} \left({\epsilon\over
1-\epsilon}\right)\left({L\over L_E}\right)^{-1} .
\label{Eq:tE}
\end{equation}
For constant values of $\epsilon$ and $L/L_E$, the growth of the black
hole from a seed mass $M_{\rm seed}$ over time $\Delta t$ is
exponential, with $M=M_{\rm seed} \exp({\Delta t/t_E})$.
Interestingly, the growth time $t_E$ has no explicit dependence on BH
mass, and so it remains the same irrespective of whether multiple
seeds grow in parallel or they coalesce to grow as a single BH.

At the high accretion rates of bright quasars, the cooling time of the
gas is shorter than its accretion time \cite{Narayan} and a thin
accretion disk with a high radiative efficiency \cite{Shakura} forms.
The existence of such disks has been confirmed recently by
microlensing observations \cite{Micro}.  The radiative efficiency of
the gas is dictated by the inner boundary of the disk at the innermost
stable circular orbit \cite{Novikov}, from where the gas plunges into
the BH and its viscous dissipation rate diminishes \cite{Shafee}. For
a non-spinning BH, $\epsilon=5.7\%$, while for a maximally-spinning
BH, $\epsilon=42\%$ \cite{Novikov,SL}.  During the prodigious growth
phase of quasars, the BH is expected to be spun-up quickly by the
infalling gas, but the orientation of the angular momentum vector of
the disk might vary considerably between different accretion
episodes. Therefore, we adopt an intermediate value between these
efficiencies in our fiducial numerical example. The fact that the
cosmic BH mass budget grew through a luminous accretion mode of a high
radiative efficiency ($\epsilon\sim 10\%$) and not through a hidden
mode of a low radiative efficiency, is demonstrated by comparing the
radiation energy density produced by quasars to the BH mass density in
the local Universe \cite{Soltan,SWM}. A high radiative efficiency is
also implied by the clustering and abundance statistics of quasars
\cite{Shankar}.

Stars form through fragmentation of cold gas in disks of galaxies
\cite{Kennicutt}.  The growth of the stellar mass budget occurs on the
dynamical timescale of the host galactic disk \footnote{The luminosity
of star forming galaxies is typically well below the Eddington limit
for their total mass.}.  The most vigorous mode of star formation
(starburst activity) is also realized in gas-rich galaxy mergers
within which cold gas is concentrated into a compact region
\cite{Cox}. The fragmentation (Jeans) mass is lowered to the scale of
stars only in environments that are denser by $\gtrsim 4$ orders of
magnitude than the mean cosmic density $\bar{\rho}$; hence, the
associated dynamical time is guaranteed to be shorter by $\gtrsim 2$
orders of magnitude than the age of the Universe [$\sim
(G\bar{\rho})^{-1/2}$] at all redshifts.  The undelayed growth in the
stellar content of galaxies through merger-driven starbursts is
ultimately limited by the global assembly rate of cold gas into
galaxies.  This rate is in turn dictated by the cosmological build-up
of dark matter halos in which gas may cool.

The deposition of energy and momentum by stars or quasars into their
gaseous environment may expel gas from the host galaxy and further
regulate the growth of the stellar and BH mass budgets. However, here
we focus on the maximum allowed growth in both components and show
that the mass density of stars could have grown faster than that of
quasar BHs at very high redshifts.  There are indications that these
maximum rates were realized in nature.  The existence of bright
quasars with BH masses $M\gtrsim 10^9M_\odot$ at $z\sim 6$ \cite{Fan},
when the age of the Universe was only $t_{\rm Hub}=10^9~{\rm
yr}[(1+z)/7]^{-3/2}$, implies that the exponential (Eddington-limited)
growth of their BHs must have persisted over a time interval $\Delta
t$ that covers much of their cosmic history \cite{Li},
\begin{equation}
{\Delta t\over t_{\rm Hub}}= 0.4 \left({1+z\over 7}\right)^{3/2}
\left({\epsilon \over 1-\epsilon}\right)\left({L\over
L_E}\right)^{-1}\ln \left({M\over M_{\rm seed}}\right).
\end{equation}
Despite this persistent growth \footnote{It is possible that the most
massive BHs at $z\sim 6$ have formed with a lower radiative efficiency
than the bulk of the BH population, allowing them to grow more rapidly
at earlier cosmic times. This might be an important selection bias,
since the quasars would not have been detected by existing surveys
\cite{Fan} if they were fainter.}, the observed quasar population did
not supply sufficient UV photons to reionize the Universe at $z>6$
even though it dominates the production of ionizing photons at
$z\lesssim 3$ \cite{Claude08}. This implies that the stellar mass
budget was able to grow faster than quasars at early cosmic times
relative to their respective growth at late times.  Below we show that
this phenomenon is a generic outcome of the evolution of structure in
the standard $\Lambda$CDM cosmology.

The production rate of ionizing photons is proportional to the growth
rate in the mass densities of BHs and stars per comoving volume in the
Universe \footnote{UV photons are produced by massive stars with
lifetimes well below the age of the Universe at the redshifts of
interest (hundreds of millions of years). Even when a massive star
radiates near the Eddington limit, its lifetime is only a few million
years because its radiative efficiency is much lower than that of a BH
(see Eq. \ref{Eq:tE}). Despite their lower efficiency, stars remain
competitive with BHs as UV sources because they generically consume a
much bigger fraction of the baryons in galaxies \cite{Geb}.}.  While
the minimum growth time of BH mass is $t_E$, the shortest growth time
of the stellar mass budget, $t_{\rm gal}$, is dictated by the assembly
rate of cold gas into galaxies. At late cosmic times, $t_E$ is much
shorter than $t_{\rm gal}$, leading to a feedback regulated mode of BH
growth in which the supply of cold gas is a limiting factor
\cite{Silk,WL03,Springel}. Below we show that at early cosmic times
the situation was reversed.

\paragraph{Mass assembly rate of galaxies.}
The minimum mass of galaxies in which BHs and stars form is dictated
by cooling considerations.  At the redshifts of interest here, the
cooling time of the gas is shorter than its dynamical time for dark
matter halos that have a virial temperature $T_{\rm vir}\gtrsim
10^4$K, above the threshold for atomic hydrogen cooling \cite{BL}.
Galaxies above this threshold are believed to have hosted the bulk of
the sources that have reionized the Universe.  Early on, the minimum
virial temperature of galaxies might have been reduced by an order of
magnitude through molecular hydrogen cooling, but ${\rm H}_2$ can be
easily dissociated by UV photons \cite{HL}. After reionization, the
minimum $T_{\rm vir}$ is expected to have increased to $\sim 10^5$K,
owing to photo-ionization heating of the intergalactic medium from
where galaxies are assembled \cite{WL06}.  The fraction of baryons
available to make stars equals the mass fraction of dark matter that
virialized in halos above the minimum $T_{\rm vir}$. We denote this
fraction by $f(z)$ and calculate the timescale for the assembly of
cold gas into galaxies from the time derivative
$\dot{f}=(df/dz)/(dt/dz)$, where $dt/dz=8.4\times 10^{7}~{\rm
yr}[(1+z)/10]^{-5/2}$, so that
\begin{equation}
t_{\rm gal}\equiv {f\over \dot{f}}\approx {3\over 2} \left|{d\ln
f\over d\ln (1+z)}\right|^{-1} t_{\rm Hub}.
\end{equation}
In the Press-Schechter formalism \cite{PS}, $f={\rm
erfc}[\delta_c(z)/\sqrt{2}\sigma(M_{\rm min})]$, where
$\delta_c=1.686/D(z)$ is the collapse threshold for an overdensity
linearly-extrapolated to $z=0$, $D(z)$ is the linear growth factor of
density perturbations (with $D(0)=1$), and $\sigma(M_{\rm min})$ is
the root-mean-square amplitude of linearly-extrapolated density
perturbations at $z=0$ on the scale of a sphere from which the minimum
galaxy mass is assembled. Refinements to the halo mass function
\cite{ST} have a negligible effect ($\lesssim 15\%$) on our results
for $t_{\rm gal}$.

\begin{figure}[th]
\includegraphics[scale=0.4]{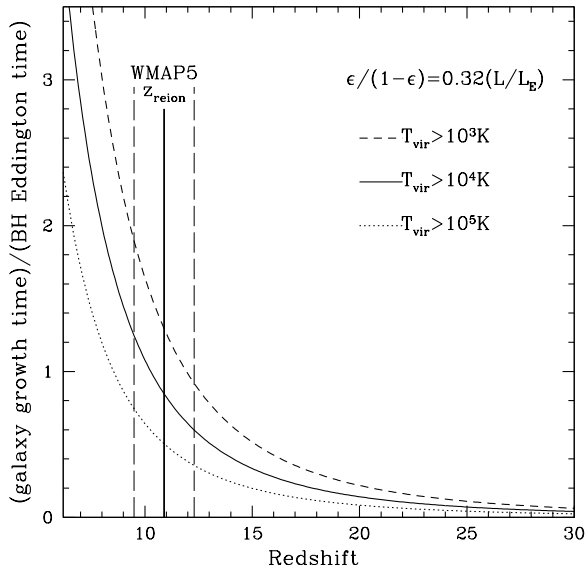}
\caption{Redshift evolution of the growth time (in $t_E$ units) for
the fraction of matter that is incorporated in dark matter halos with
virial temperatures above $10^3$K (dashed line), $10^4$K (solid), and
$10^5$K (dotted). Conservatively, we adopt a value of
$[\epsilon/(1-\epsilon)]=0.32(L/L_E)$, which is a factor of $\sim 2$
smaller than the lower limit implied by observations of high redshift
quasars \cite{Shankar}.  The vertical lines correspond to the central
value (solid line) plus or minus one standard deviation (dashed lines)
for the redshift of reionization, $z_{\rm reion}$, based on the WMAP5
data \cite{Komatsu}.  Prior to reionization, the assembly of cold gas
into galaxies occurs at a faster rate than the Eddington-limited
growth of BHs. }
\end{figure}

\paragraph{Results.}
Figure 1 shows the ratio $(t_{\rm gal}/t_E)$ as a function of redshift
$z$ for the concordance $\Lambda$CDM cosmology \cite{Komatsu}.  The
galaxy growth time is calculated as the fraction of matter that is
incorporated in dark matter halos with virial temperatures above
$10^3$K (dashed line), $10^4$K (solid), and $10^5$K (dotted). We adopt
$[\epsilon/(1-\epsilon)]=32\%(L/L_E)$, corresponding to a BH
spin-averaged radiative efficiency of $\epsilon=(6+42)/2=24\%$~ for
$L=L_E$ or to $\epsilon=5.7\%$ for the characteristic value \cite{Kol}
of $L/L_E\approx 0.2$. This choice is conservatively smaller by a
factor of $\sim 2$ than the lower limit implied by observational data
on high-redshift quasars \cite{Shankar}; a corresponding increase in
this value by some factor would have lowered the plotted curves by the
same factor and strengthened our conclusions.  The three vertical
lines mark the central value (solid line) plus or minus one standard
deviation (68\% confidence; dashed lines) for the redshift of
instantaneous reionization, as inferred from the WMAP5 data
\cite{Komatsu}.  Prior to reionization, the assembly of cold gas in
galaxies occurs at a faster rate than the Eddington-limited growth of
BHs for all three threshold values of $T_{\rm vir}$.

At early cosmic times, galaxies are on the exponential tail of the
Press-Schechter mass function, and the mass of cold gas in them grows
faster than $t_E$. BH growth through gas accretion is not able to track
this early rapid growth.  At late times, quasar growth is limited by
the much slower rate at which fresh cold gas is infused into the
centers of galaxies through episodic mergers.  For the
post-reionization case of $T_{\rm vir}>10^5$K, we find that
$t_E\lesssim 0.5 t_{\rm gal}$ only at $z\lesssim 6$.  It takes several
$e$--folding times for quasars to build-up their mass density and
catch up with the UV production by star formation in common galaxies
at lower redshifts.  Despite the initial delay in their growth,
quasars eventually dominate the production rate of ionizing photons by
$z\sim 3$ \cite{Claude08}. Within 1--2 billion years after the big
bang, the growth in the cosmic mass density of BHs starts to be
regulated by feedback on their gaseous environment (owing to their
increased mass) and by the depletion of their cold gas reservoir
(aided by the declining merger rate of galaxies). The characteristic
time for doubling the mass of galaxies traces the age of the Universe
($\sim 2.3\times 10^9~{\rm yr}[(1+z)/4]^{-3/2}$), and keeps increasing
relative to $t_E$ with decreasing redshift. The mode of
feedback-regulated growth \cite{Silk,WL03,Springel} becomes critical
at these low redshifts.

During the early history, when BH growth is not yet regulated by
feedback or by the exhaustion of the cold gas reservoir in galactic
nuclei, the comoving mass density of accreting BHs ($\rho_{\rm BH}$)
grows at a rate,
\begin{equation}
\dot{\rho}_{\rm BH}= {\rho_{\rm BH}/t_E} + \dot{\rho}_{\rm seed},
\label{coevolution}
\end{equation}
where $\dot{\rho}_{\rm seed}$ is the formation rate density of BH
seeds.  If the BH seeds cannot ionize the Universe on their own (i.e.,
if reionization did not result from the accretion luminosity of
stellar-mass BHs), then Fig. 1 shows that stars are required and able
\footnote{Aside from the Eddington limit, the growth rate of the early
BH population could have been inhibited by gravitational wave recoil
of merger remnants out of the shallow potential wells of the first
dwarf galaxies \cite{Tanaka,Oleary}.} to make reionization happen by
$z=10.9\pm 1.4$.

At late times, the BH mass density grows exponentially until feedback
and the consumption of cold gas (which were not included in
Eq. \ref{coevolution}) start to regulate its growth \footnote{This
regulation would lead to a saturation value of $\rho_{\rm BH}$ that is
proportional to the comoving density of stars $\rho_\star$ at very
late times, as indicated by observations \cite{Magorrian}.}.  By
requiring that quasars match the UV production rate per comoving
volume of stars only as late as $z\sim 6$--$3$ \cite{Claude08}, we
infer that they fall short of matching it at $z\sim 11$ by orders of
magnitude. The transition redshift between the early domination by
stars and the late domination by quasars is sensitive to uncertain
parameters (which may also be redshift dependent), such as the unknown
feedback strength and the radiative efficiency of these source
populations.

The radiative efficiency of stars and BHs is obviously different.  The
total number of emitted ionizing photons per baryon incorporated into
stars ranges \cite{BKL} between $4\times 10^3$ [for a present-day (Pop
II) mass function] and $\sim 10^5$ [for metal-free massive stars (Pop
III)]. Supermassive BHs produce $\sim 2\times 10^7 \epsilon$ ionizing
photons per accreted baryon \cite{Telf}. The observation that quasars
dominate the cosmic UV production rate only at low redshifts
\cite{Claude08} implies that their BH formation efficiency out of cold
gas in galaxies is lower than that of stars by several orders of
magnitude.  This is likely caused by angular momentum which
distributes the cold galactic gas in a large-scale disk that fragments
into stars long before tidal or viscous torques enable a small
fraction of this gas to feed the central BH.  The inference of a low
BH formation efficiency is confirmed by data on the cumulative mass
budgets of supermassive BHs and stars in the local Universe
\cite{Magorrian}.

The fundamental limitation presented by Fig. 1 applies also to an
early population of stellar-mass black holes, because those would
still require extensive accretion in order to reionize the
Universe. The comoving luminosity density of any BH population is
simply proportional to its cumulative accretion rate, and the
Eddington limitation on the accretion time is independent of the mass
distribution of the early BHs.  Indeed, the smallness of the unresolved
component of the soft X-ray background places severe constraints on
the cumulative mass density of an early population of accreting BHs
irrespective of its mass distribution \cite{Dijk}.

\paragraph{Observable signatures.}
The delay associated with the Eddington-limited growth of black holes
in the early Universe can be probed through a number of observational
methods. First, gravitational wave signals from coalescing BH binaries
at high redshifts -- which are detectable by {\it LISA} 
\footnote{http://lisa.nasa.gov/} and possibly also by Advanced LIGO
\cite{WL04}, can be used to probe the mass function of massive BHs as
a function of redshift.  Second, future X-ray missions such as the
{\it International X-ray Observatory}
\footnote{http://ixo.gsfc.nasa.gov/}, and infrared telescopes such as
{\it JWST} \footnote{http://www.stsci.edu/jwst/}, or new ground-based
instruments, may detect fainter quasars at higher redshifts than those
accessible with present-day telescopes. Third, BHs that escaped from
their dwarf galaxies by gravitational-wave recoil at early cosmic
times, are potentially detectable in the Milky-Way halo through the
compact star clusters that surround them \cite{OL}. Finally, extending
current simulations of quasar growth \cite{Springel,Li} to the higher
redshifts of reionization ($z\gtrsim 12$) would refine predictions for
the ionized bubble sizes, which may be probed by future 21cm
observatories \cite{Telf,Loeb08}.

\bigskip
\bigskip

\paragraph*{Acknowledgments.}
I thank J. Pritchard for helpful comments.  This work was supported in
part by Harvard University funds.

\end{document}